\documentclass[aip,jcp]{revtex4-1}

\usepackage{graphicx}
\usepackage{float}
\usepackage{amsmath}

\begin{document}

\title{Incorporating multi-body effects in SAFT by improving the description of the reference system. I.  Mean activity correction for 
 cluster integrals in the reference system}
\author{Artee Bansal}
\author{D.\ Asthagiri}
\author{Kenneth R. Cox}
\author{Walter G. Chapman}\thanks{wgchap@rice.edu}
\affiliation{Department of Chemical and Biomolecular Engineering, Rice University, Houston}
\date{\today}
\vfill

\begin{abstract}
	
A system of patchy colloidal particles interacting with a solute that can associate multiple times in any direction is a useful model for patchy colloidal mixtures. Despite the simplicity of the interaction, because of the presence of multi-body correlations 
predicting the thermodynamics of such systems remains a challenge. Earlier Marshall and Chapman developed 
a multi-body formulation for such systems wherein the cluster partition function for the hard-sphere solvent molecules 
in a defined inner-shell (or coordination volume) of the hard-sphere solute is used as the reference within the statistical association 
fluid theory formalism.  The multi-body contribution to these partition functions are obtained by ignoring the bulk solvent, thus limiting the applicability of the 
theory to low system densities.  Deriving inspiration from the quasichemical theory of solutions where these partition functions occur in the guise of 
equilibrium constants for cluster formation, we develop a way to account for the multi-body correlations including the effect of the bulk solvent. 
We obtain the free energy to evacuate the inner-shell, the chemistry contribution within quasichemical theory, from simulations of the hard-sphere reference. This chemistry contribution reflects association in the reference in the presence of the bulk medium. The gas-phase partition functions are then augmented by a mean activity factor that is adjusted to reproduce the chemistry contribution. We show that the updated partition function provides a revised reference that better captures the 
distribution of solvent around the solute up to high system densities. Using this updated reference, we find that theory better describes both 
the bonding state and the excess chemical potential of the colloid in the physical system.

\end{abstract}

\maketitle

\section{Introduction}
The physical mechanisms governing the structure, thermodynamics, and dynamics of particles interacting with short-range anisotropic interactions are
of fundamental scientific interest in the quest to understand how inter-molecular interactions dictate macroscopic structural and functional organization   \cite{jackson_phase_1988,marshall_thermodynamic_2014,marshall_cluster_2014,russo_re-entrant_2011,tavares_criticality_2009}. Patchy colloids, particles with engineered directional interactions, are archetypes of such systems, with numerous emerging applications in designing materials from the nanoscale level \cite{glotzer_anisotropy_2007,pawar_fabrication_2010,bianchi_patchy_2011,sciortino_gel-forming_2008,cordier_self-healing_2008}. 
Experiments on patchy colloidal systems have focused on the synthesis of different kinds of self assembling units and their
consequence for the emergent structure \cite{zhang_patterning_2005,yake_site-specific_2007,snyder_nanoscale_2005,chen_directed_2011,pawar_patchy_2008,wang_colloids_2012,romano_colloidal_2011,yi_recent_2013}.  
Complementing these experimental studies, molecular simulations have also sought to understand how the anisotropy of interactions 
determines the emergent structure \cite{zhang_self-assembly_2004,zhang_self-assembly_2005,coluzza_design_2012,de_michele_dynamics_2006} and
the phase behavior and regions of stability in the phase diagram\cite{bianchi_phase_2006,bianchi_theoretical_2008,foffi_possibility_2007,giacometti_effects_2010,liu_vapor-liquid_2007,romano_gasliquid_2007}.  But despite the simplicity in describing and engineering the inter-molecular interactions, a general theory to predict the phase behavior is not yet available. The present article 
develops a multi-body theory that is a step towards developing a comprehensive theory of such colloidal mixtures.

Wertheim's perturbation theory in the form of statistical associating fluid theory (SAFT) \cite{jackson_phase_1988,wertheim_fluids_1984,chapman_new_1990,wertheim_fluids_1984-1,heras_phase_2011,heras_bicontinuous_2012} has proven to be an effective framework in describing systems with short range directional interactions and is thus of natural interest in describing patchy colloids \cite{bianchi_theoretical_2008,liu_vapor-liquid_2009,sciortino_self-assembly_2007}. In Wertheim's approach the association interaction is a perturbation from a non-associating reference fluid (typically a hard-sphere or Lennard-Jones fluid). The association contribution is obtained by equating the unbonded pair correlation function to the reference pair-correlation function and it is implicitly assumed that each associating site can only bond once. 
However, for a spherically symmetric colloid, the single bonding condition does not hold. Moreover, the 
pair correlation information is not enough, especially for a dense fluid, to model the multi-body effects in either the physical system or the hard-sphere
reference.  

Several recent studies acknowledge the importance of multi-body effects. In atomistic simulation studies of a patchy colloidal mixture, Liu et al.\ \cite{liu_vapor-liquid_2007} recognized that patchiness broadens the vapor-liquid coexistence curve over that for a system with isotropic interactions. To account for this, they incorporated a square well reference (instead of the usual hard-sphere reference) in Wertheim's first order perturbation theory. Using this approach they could describe qualitatively the increasing critical temperature with increasing number of patches, but the quantitative agreement with simulations was limited \cite{liu_vapor-liquid_2009}.  Kalyuzhnyi and Stell \cite{kalyuzhnyi_effects_1993} reformulated Wertheim's multi-density formalism \cite{wertheim_fluids_1986} in integral equation approach to incorporate spherically symmetric interactions but the solution becomes complex for large values of bonding states. Key extensions to Wertheim's theory were provided by Marshall and Chapman   \cite{marshall_wertheims_2012,marshall_thermodynamic_2014,marshall_resummed_2014} for multiple bonding per site and cooperative hydrogen bonding.

To incorporate multi-body effects in SAFT when the association potential of the solute is assumed to be spherically symmetric, as opposed to directional, Marshall and Chapman \cite{marshall_molecular_2013,marshall_thermodynamic_2013} developed a new theory beyond Wertheim's multi-density formalism for multi-site associating fluids \cite{wertheim_fluids_1986}. The theory requires the multibody correlation function for solvent around the solute in a non-associating reference fluid. For the reference fluid, the multi-body correlations were approximated by cluster partition functions in isolated clusters and by application 
of linear superposition of the pair correlation function. This approximation works well for low densities of the system, but the 
higher order correlations become important at higher solvent densities. To accurately incorporate multi-body effects in the SAFT framework, a better representation of multi-body correlations in the hard sphere reference is thus required.

Here we build on the earlier work by Marshall and Chapman \cite{marshall_molecular_2013,marshall_thermodynamic_2013}. The spherically symmetric and patchy colloids are modeled as hard spheres of equal diameter ($\sigma$) and short range association sites. Recognizing that using the gas-phase cluster partition functions is akin to the primitive quasichemical approximation, we derive inspiration from developments in the quasichemical theory \cite{pratt_quasichemical_2001,pratt_selfconsistent_2003} to better account for the role of the bulk material in modulating the clustering of the reference solvent around the reference solute. We then use the improved reference
within the multi-density formalism \cite{wertheim_fluids_1986,marshall_molecular_2013,marshall_thermodynamic_2013}. 

The rest of the paper is organized in the following way.  In Section~\ref{sc:bentheory} we discuss the Marshall-Chapman\cite{marshall_thermodynamic_2013} theory and highlight the need for improvement suggested by comparing the results of theory with Monte Carlo simulations. In Section~\ref{sc:qct} we present elements of the quasichemical approach and discuss how it can be used to provide an updated reference, and in Section~\ref{sc:correction} the theory incorporating the updated reference is presented. We present the results in Section~\ref{sc:results}.

\section{Theory} 
   \subsection{Mixtures with spherically symmetric and directional association potential }\label{sc:bentheory}
\begin{figure}[h!]
\includegraphics[scale=0.95]{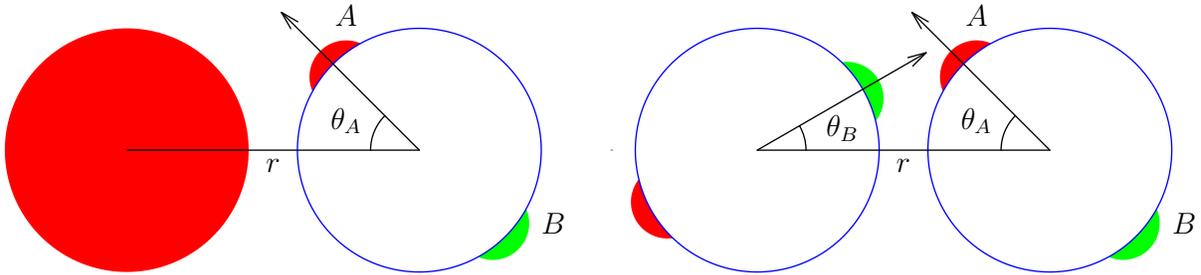}
 \caption{Association between solute and solvent (left) and solvent molecules (right). $r$ is the center-to-center distance and $\theta_A$ and $\theta_B$ are the orientation of the attractive patches $A$ and $B$ relative to line connecting the centers. Note the spherical solute (colored red) can only interact with the patch $A$ (colored red).}\label{fig:passo}    
\end{figure}

Consider a mixture of solvent molecules, $p$, with two directional sites (labeled $A$ and $B$) and spherically symmetrical, $s$, solute molecules.
For solvent-solvent association, only bonding between $A$ and $B$ is allowed and  the size of sites  is such that single bonding condition holds (Fig.~\ref{fig:passo}). The solute molecule can bond with site $A$ of the solvent; the isotropic attraction ensures the solute can bond multiple solvent molecules (Fig.~\ref{fig:passo}). In the infinitely dilute regime considered here, we ignore the
association between the solutes themselves. 

   The association potential for solvent-solvent  $(p,p)$  and solute-solvent $(s,p)$ molecules is given by:
   \begin{equation}
   u_{AB}^{(p,p)}{(r)}=
   \begin{cases}
   -\epsilon_{AB}^{(p,p)}, r<r_c \,\text{and}\, \theta_A\leq \theta_c^{(A)}\,\text{and}\,\theta_B \leq \theta_c^{(B)}
   \\
   0   \text{ \ \ \ \ \ otherwise}
   \\    
   \end{cases}
   \label{eq:1}
   \end{equation}
   
   \begin{equation}
   u_A^{(s,p)}{(r)}=
   \begin{cases}
   -\epsilon_A^{(s,p)}, r<r_c\, \text{and}\, \theta_A\leq \theta_c^{(A)}
   \\
   0   \text{ \ \ \ \ \ otherwise}
   \\    
   \end{cases}
   \label{eq:2}
   \end{equation}
   where subscripts $A$ and $B$ represent the type of site and $\epsilon$  is the association energy. $r$ is the distance between the particles and $\theta_A$ is the angle between the vector connecting the centers of two molecules and the vector connecting association site $A$ to the center of that molecule (Fig.~\ref{fig:passo}).  The critical distance beyond which particles do not interact is $r_c$ and $\theta_c$ is the solid angle beyond which sites cannot bond. 
   
 The role of attractions between solvent, $p$, molecules is accounted by standard first order thermodynamic perturbation theory \cite{jackson_phase_1988} (TPT1). For the association contribution to intermolecular interactions between spherically symmetric solute ($s$ molecules) and solvent ( $p$ molecules) with directional sites, Marshall and Chapman \cite{marshall_molecular_2013,marshall_thermodynamic_2013} developed a theory based on generalization of Wertheim's single chain approximation\cite{wertheim_fluids_1984-1,wertheim_fluids_1986}. By including graph sums for all the possible arrangements of the solvent around the solute (i.e.\ one solvent around solute, two solvents around solute, etc.), Marshall and Chapman obtained the free energy expression for the mixture as: 
     \begin{eqnarray}
     \frac{{{A^{AS}}}}{{NkT}} & = & {x^{(s)}}( {\ln X_o^{( s )} + \frac{{\bar n}}{2}})  +  ( {1 - {x^{( s )}}} )\sum\limits_{A \in {\Gamma ^{( p )}}} {( {\ln X_A^{( p )} - \frac{{X_A^{( p )}}}{2} + \frac{1}{2}} )} 
     \end{eqnarray}
    where the superscript $(s)$ and $(p)$ indicates the molecule type; $x^{(s)}$ is the mole fraction of $s$ molecules in the mixture; $X_o^{(s)}$  is the monomer fraction of $s$ molecules, i.e.\ it is the fraction of $s$ molecules that are not bonded. $\Gamma^{(p)}={(A,B)}$ is the set of attractive sites on the $p$ molecules, and 
  $X_A^{( p )}$and $X_B^{( p )}$ are the fraction of $p$ molecules not bonded at patch $A$ and $B$, respectively.  The average number of bonds 
per $s$ particle is given by $\bar n$ and this quantity can be obtained from, $\rho_n^{(s)}$, the density of $s$ particles bonded to $n$ number of solvent molecules.
$\rho_n^{(s)}$ is given by
      \begin{eqnarray}
      \rho _n^{(s)} = \frac{\Delta c_n^{(0)}}{V}
      \label{eq:29}
      \end{eqnarray}
 where
\begin{widetext}         
  \begin{eqnarray}
  	\Delta c_n^{\left( 0 \right)} & = & \frac{\rho _0^{(s)} {\left( {\rho ^{(p)}X_A} \right)}^n} {\tilde \Omega^{n + 1} n!} \int d(1)....d(n + 1) \,
  	g_{HS}( 1....n + 1) \cdot \prod\limits_{k = 2}^{n + 1} {\left( f_{as}^{(s,p)} ( 1,k) \right)}  \, .
  	\label{eq:14}
  \end{eqnarray}
\end{widetext} 
In Eq.~\ref{eq:14}, $\rho_0$ is the monomer density, $\rho^{(p)}X_A$ is the density of patchy molecules not bonded at site A , and $\tilde \Omega (=4\pi)$ is the total number of orientations. The many body correlation for the hard sphere reference fluid, $g_{HS}(1...n+1)$, can be represented in terms of the 
hard sphere cavity correlation function as
   \begin{equation}
   {g_{HS}}\left( {1....n + 1} \right) = {y_{HS}}\left( {1....n + 1} \right)\prod\limits_{\{ l,k\} } {{e_{HS}} \left( {r{}_{lk}} \right)} \, 
   \label{eq:15}
   \end{equation}
 where  $e_{HS}(r_{lk}) = \exp(-u_{HS}/kT) $ are reference system $e$-bonds which serve to prevent hard sphere overlap in the cluster; 
 $e_{HS}(r_{lk}) = 0$ for $r_{lk} < \sigma$. Marshall and Chapman \cite{marshall_molecular_2013} approximated these many body cavity correlation functions with first order superposition of  pair cavity correlation function at contact times
 a second order correction ($\delta ^{\left( n \right)}$)
   \begin{equation}
   {y_{HS}}\left( {1....n + 1} \right) \approx y_{HS}^n\left( \sigma \right){\delta ^{(n)}} \, .
    \label{eq:yhs}
   \end{equation}
As is usual in SAFT, the contribution due to association is given by an averaged $f$-bond and factored
outside the integral. Then the integral, with positions in spherical coordinate system, in Eq.~\ref{eq:14} becomes
\begin{widetext}
   \begin{eqnarray}
 {\Xi ^{(n)}}  = \prod\limits_{k = 2}^{n + 1} {\int\limits_0^{2\pi } {\int\limits_{ - 1}^1 {\int\limits_\sigma^{{r_c}} {d{\phi _{1,k}}d\cos {\theta _{1,k}}d{r_{1,k}}r_{1,k}^2} } } } {\prod\limits_{j > i = 1}^n {e(i,j)} } 
    \label{eq:110}
   \end{eqnarray}
   \end{widetext}
   which gives the cluster partition function for an isolated cluster with $n$ solvent hard spheres around a hard sphere solute 
   in the volume defined by the hard sphere diameter $\sigma$ and $r_c$. These partition functions can be obtained as
   \begin{equation}
   {\Xi ^{( n )}} = \nu _b^n{P^{( n )}}
   \label{eq:11}
   \end{equation}
   where $\nu _b$  is the bonding volume and $P^{( n )}$ is the probability that there is no hard sphere overlap  for randomly generated $p$ molecules in the bonding volume (or inner-shell) of $s$ molecules. A hit-or-miss Monte Carlo \cite{hammersley,pratt_quasichemical_2001} approach to calculate $P^{(n)}$ proves inaccurate for large values of $n$ ( $n>8$). But since 
   \begin{equation}
   {P^{( n)}} = P_{insert}^{( n )}{P^{( {n - 1} )}} \, , 
   \label{eq:12}
   \end{equation}
 where $P_{insert}^{(n)}$  is the probability of inserting a {\em{single}} particle given $n-1$ particles are already in the bonding volume, an 
 iterative procedure can be used to build the higher-order partition function from lower order one \cite{marshall_molecular_2013}.   The one-particle insertion probability $P_{insert}^{(n)}$ is easily evaluated using hit-or-miss Monte Carlo.  The maximum number of $p$ molecules for which a non-zero insertion probability can be obtained defines $n^{max}$. 

Eq.~\ref{eq:14} reduces to
  \begin{equation}
  	\frac{\Delta c_n^{(0)}}{V} = \frac{1}{{n!}}\rho _0^{(s)} \Delta^n {\Xi ^{(n)}}{\delta ^{(n)}}   \,  .
  	\label{eq:281}
  \end{equation} 
with the potential defined by Eq.~\ref{eq:2} and approximation Eq.~\ref{eq:yhs}. The fraction of spherically symmetric molecules bonded $n$ times is 
  \begin{equation}
  	X_n^{\left( s \right)} = \frac{{\frac{1}{{n!}}{\Delta ^n}{\Xi ^{\left( n \right)}}{\delta ^{\left( n \right)}}}}{{1 + \sum\limits_{n = 1}^{{n^{\max }}} {\frac{1}{{n!}}{\Delta ^n}{\Xi ^{\left( n \right)}}{\delta ^{\left( n \right)}}} }}{{,\quad       n > 0}} \, , 
  	\label{eq:7}
  \end{equation}
and the fraction bonded zero times is
  \begin{equation}
  	X_0^{\left( s \right)} = \frac{1}{{1 + \sum\limits_{n = 1}^{{n^{\max }}} {\frac{1}{{n!}}{\Delta ^n}{\Xi ^{\left( n \right)}}{\delta ^{\left( n \right)}}} }} \, .
  	\label{eq:8}
  \end{equation}
For a two patch solvent, 
   \begin{equation}
   \Delta  = {y_{HS}}\left( \sigma \right)X_A^{\left( p \right)}{\rho ^{\left( p \right)}}f_A^{\left( {s,p} \right)}\sqrt {{\kappa _{AA}}} \, .
   \label{eq:9}
   \end{equation}
$\kappa_{AA}$ is the probability that molecule $p$ is oriented such that patch $A$ on $p$ bonds to $s$;  
$f_A^{\left( {s,p} \right)}$  is the  Mayer function for association between $p$ and $s$ molecules 
  \begin{equation}
   f_A^{(s,p)} = \exp (\varepsilon_A^{(s,p)}/kT) - 1 \, .
  \label{eq:10}
   \end{equation}
Finally, the average number of patchy colloids associated with a spherically symmetric colloid is given by: 
   \begin{equation}
   \bar n = \sum\limits_n {n{X_n}}  \, ,
   	\label{eq:81}
   \end{equation}

The fraction of solvent not bonded at site $A$ and site $B$ can be obtained by simultaneous solution of the following equations:
     
     \begin{equation}
     X_A^{\left( p \right)} = \frac{1}{{1 + \xi {\kappa _{AB}}f_{AB}^{\left( {p,p} \right)}{\rho ^{\left( p \right)}}X_B^{(p)} + \frac{{{\rho ^{\left( s \right)}}}}{{{\rho ^{\left( p \right)}}}}\frac{{\overline n }}{{X_A^{(p)}}}}}  \, ,
     \end{equation}
     
     \begin{equation}
     X_B^{\left( p \right)} = \frac{1}{{1 + \xi {\kappa _{AB}}f_{AB}^{\left( {p,p} \right)}{\rho ^{\left( p \right)}}X_A^{(p)}}} \, .
     \end{equation}
  where   
     \begin{equation}
     \xi  = 4\pi {d^2}\left( {{r_c} - \sigma} \right){y_{HS}}\left( \sigma \right) \,
     \end{equation}  
     \begin{equation}
     f_{AB}^{\left( {p,p} \right)} = \exp ( \varepsilon _{AB}^{({p,p})}/kT)-1 \, .
     \end{equation}

As will be seen below,  the above approach works very well for low solvent densities ($\rho\sigma^3 \leq 0.6$). However, 
as can be intuitively expected, the approximation of using a gas-phase cluster partition function (Eq.~\ref{eq:110}) is less accurate at 
higher solvent densities that are of practical interest in modeling a dense solvent. (The approximation embodied
in Eq.~\ref{eq:yhs} and in factoring the association contributions outside the integral are likely of less concern given
the very short-range of attractions relative to the size of the particle.) Borrowing ideas from quasi-chemical theory,
we next consider how to better approximate the reference cluster partition function.

    \subsection{Quasi-chemical theory for solvation of hard-core solutes}\label{sc:qct}  

Consider the equilibrium clustering reaction within some defined coordination volume of the solute $A$ in a bath of solvent $S$ molecules
    \begin{equation}
    A{S_{n = 0}} + {S_n} \rightleftharpoons A{S_n} \, . 
    \label{eq:20}
    \end{equation}
The the equilibrium constant  is
    \begin{equation}
    {K_n} = \frac{{{\rho _{A{S_n}}}}}{{{\rho _{A{S_{n = 0}}}}\rho _s^n}} \, ,
    \label{eq:21}
    \end{equation}
where $\rho_{AS_n}$ is the density of  species $AS_n$ and $\rho_s$ is the density of the solvent. A mass balance
then gives the fraction of $n$-coordinated solute as    
\begin{equation}
     {p_n} = \frac{{{K_n}^{}\rho _s^n}}{{1 + \sum\limits_{m \ge 1} {{K_m}^{}\rho _s^m} }} \, .
      \label{eq:16}
     \end{equation}
The $n=0$ term, $p_0$, is of special interest: $\ln p_0 = -\ln (1 + \sum\limits_{m \ge 1} K_m \rho _s^m) $ is free energy of allowing solvent molecules to populate a formerly empty coordination shell.  Observe that the $\ln p_0$ expansion is determined by the various coordination states.  
 In the language of quasichemical theory, $\ln p_0$ is called the chemical term \cite{lrp:book,lrp:cpms,merchant_thermodynamically_2009}. Because the bulk medium pushes solvent into the coordination volume, an effective attraction exists between the solute and solvent even 
for a hard-sphere reference. 

In the primitive quasichemical approximation \cite{merchant:jcp11b}, the equilibrium constants are evaluated by neglecting the effect of the bulk medium, i.e.\ for an isolated cluster. Thus $K_n \approx K_n^{(0)}$ \cite{pratt_quasichemical_2001}, where 
 \begin{equation}
  n!K_n^{(0)} = \int\limits_A {d{{\vec r}_1} \cdots  \int\limits_A d{{\vec r}_n} {\prod\limits_{j > i = 1}^n {e(i,j)} } } \,
 \label{eq:210}
\end{equation}
where $A$ in the integral indicates that the integration is restricted to the defined coordination volume.  Comparing Eqs.~\ref{eq:110} and \ref{eq:210}, 
clearly  $n! {K_n}^{(0)} \equiv \Xi ^{(n)}$, establishing a physical meaning for Eq.~\ref{eq:110}. 
          
It is known that the primitive approximation leading to Eq.~\ref{eq:210} introduces errors in the estimation of $\ln p_0$ \cite{pratt_quasichemical_2001,pratt_selfconsistent_2003}, especially for systems where the interaction of the solute with the solvent is not sufficiently stronger than the interaction amongst solvent particles \cite{merchant:jcp11b}. For hard spheres we must then expect the primitive approximation to fail outside the limit of low 
solvent densities.

One approach to improve the primitive approximation is to include an activity coefficient $\zeta$, such that the predicted occupancy
in the observation volume is equal to occupancy, $\langle n\rangle$, expected in the dense reference \cite{pratt_quasichemical_2001}
        \begin{eqnarray}
        \sum\limits_n {n{K_n}^{\left( 0 \right)}\rho _S^n{\zeta ^n}}  = \left\langle n \right\rangle \sum\limits_n {{K_n}^{\left( 0 \right)}\rho _S^n{\zeta ^n}} \, .
         \label{eq:22}
        \end{eqnarray}
Here the factor $\zeta$ functions as a Lagrange multiplier to enforce the required occupancy constraint ($\langle n\rangle$). Physically, 
$\zeta$ is an activity coefficient that serves to augment the solvent density in the observation volume over that predicted by the gas-phase
equilibrium constant $K_n^{(0)}$. In principle, $\zeta$ should itself be $n$-dependent, but here we assume a mean-activity value that is the same for all $n$ for the given density. 
With the above consistency requirement, $p_0$ becomes 
         \begin{equation}
        {p_0} = \frac{1}{{1 + \sum\limits_{m \ge 1} {{K_m}^{(0)}{\zeta ^m}\rho _S^m} }} \, .
         \label{eq:23}
         \end{equation}

In the original implementation of the above idea for the problem of forming a cavity in a hard-sphere fluid, the consistency condition was the average occupancy of the cavity, a quantity that is known given the density of the liquid \cite{pratt_quasichemical_2001}.  
While this constraint improves upon the primitive approximation, for high densities this approximation does not predict the correct free energy to open a cavity in the hard-sphere liquid. In a subsequent work \cite{pratt_selfconsistent_2003}, in addition to $\zeta$ a solvent coordinate-dependent molecular field  was introduced to enforce the required uniformity of density  (for a homogeneous, isotropic system) inside the observation volume. With the molecular field, the predicted free energy to create a cavity in the fluid was found to be in excellent agreement with the Carnahan-Starling \cite{mansoori_equilibrium_1971} result up to high 
densities.   Both these approaches seek to predict hard-sphere properties from few-body information. However, here we acknowledge the
availability of extensive simulation data on hard-spheres, and thus seek a $\zeta$ that will reproduce the free energy to evacuate the inner-shell around the reference solute (Eq.~\ref{eq:23}), indicated as ``Theory + $p_0$ constraint" in figures below. Additionally, we also tested a $\zeta$ that enforces Eq.~\ref{eq:22}, indicated as
``Theory + $\langle n \rangle$ constraint" in figures below. 

 Using the above ideas from QCT, we obtain corrections to be applied in the original Marshall-Chapman \cite{marshall_molecular_2013} theory.  Note that 
 the probability of having centers of $n$ solvent molecules inside the observation shell of the solute at the origin (0) is
 \begin{eqnarray}
 {P^{\left( n \right)}} & = &\frac{{{\rho ^n}}}{{n!}}\int\limits_{states} {d{{\vec r}_1} \cdots d{{\vec r}_n}{g_{HS}}\left( {0,{{\vec r}_1} \cdots {{\vec r}_n}} \right)} \nonumber \\
 & = & \frac{{{\rho ^n}}}{{n!}}{\left\langle {{g_{HS}}\left( {0,{{\vec r}_1} \cdots {{\vec r}_n}} \right)} \right\rangle _{states}}{\Xi ^{\left( n \right)}}
 \end{eqnarray}
Thus instead of Eq.~\ref{eq:yhs}, we will approximate the cavity correlation function by
 \begin{equation}
 {\left\langle {{y_{HS}}\left( {0,{{\vec r}_1} \cdots {{\vec r}_n}} \right)} \right\rangle _{states}} \approx {\zeta ^n} \, .
 \label{eq:250}
 \end{equation}

\section{Methods}
  
\subsection{Monte Carlo simulation of associating system}   
     MC simulations were performed for the associating mixtures to test the theory against simulation results. The associating mixture contains one solute with spherically symmetric site and solvent molecules with 2 sites. The association energy $\epsilon=7$~$k_BT$ was used for all pair-wise associations, where $k_{\rm B}$ is the
     Boltzmann constant and $T$ the temperature. (In simulations we set $T=298.15$~K.)  The patchy-solvent particles can bond provided the center-to-center distance $r \leq 1.1\sigma$ and the 
 angle between the A site of one particle and B site on another satisfies $\theta \leq 27^\circ$.  The solvent can associate with
 the solute provided $r \leq 1.1\sigma$ and the angle between the A site and the line connecting the
center of the solute and solvent is below $27^\circ$. All simulations comprise 863 solvent particles and 1 solute. 

 The excess chemical potential of coupling the colloid with the solvent was obtained using thermodynamic
 integration, 
 \begin{eqnarray}
\beta \mu^{Asso} = \epsilon \int_0^1 \langle \beta \psi \rangle_{\epsilon\cdot \lambda} d\lambda \, 
\end{eqnarray}
where $\langle \beta \psi \rangle_{\epsilon.\lambda}$ is the average binding energy of solute with the solvent with the solute-solvent
interaction strength $\lambda$ and $\beta = 1 / k_BT$. The integration was performed using a three-point Gauss-Legendre quadrature \cite{Hummer:jcp96}. At each
coupling strength, the system was equilibrated over 1~million sweeps, where a sweep is an attempted move for every particle. 
The translation/rotation factor was chosen to yield an acceptance ratio between $0.3-0.4$. These parameters were kept constant
in the production phase which also extended for 1~million sweeps. Binding strength data was collected every 100 sweeps for analysis. Statistical uncertainty  in $\mu^{Asso} $ was obtained using the Friedberg-Cameron approach \cite{allen:error,friedberg:1970}. 
     
Besides $\mu^{Asso}$, it is of significant interest to compare the predictions of the bonding state of the colloid ($X$) with 
simulations. $X$ is the number of times the solute is bonded. Given $p_n$, the probability of observing $n$ solvent in the coordination 
volume, $P(X)$, the probability of observing the $X$-bonded state is given by 
 \begin{equation}
P(X) = \sum p_n P(X|n) \, ,
\label{eq:bayes}
\end{equation}      
where $P(X|n)$ is the probability of observing an $X$ bonded state of the solute given $n$ solvent particles are in the 
coordination volume. Of course, $X \leq n$. 

To better reveal these low-$X$ states, we used an ensemble reweighing approach\cite{merchant_water_2011}. Essentially biases (calculated
iteratively) are used to sample $n$ as uniformly as possible. The distribution $\{p_n\}$ is readily obtained from the reweighed
probabilities $\{\bar{p}_n\}$ and the biases. For each $n$ in the biased simulation, the distribution of $X$ is obtained and $P(X|n)$ constructed. Then from Eq.~\ref{eq:bayes}, the $P(X)$ distribution is composed. For these simulations, as above, the system
was equilibrated over 1~million sweeps and data collected over a production phase of 1~million sweeps. 

     \subsection{Cluster partition function}
We recapitulate the calculation of $P_{insert}^n$ (Eq.~\ref{eq:12}) presented earlier in Refs.~\onlinecite{marshall_molecular_2013,marshall_thermodynamic_2013}. For $P_{insert}^n$, with the solute hard sphere at the center of coordinate system, the trial position of the particle in the coordination volume is randomly generated. The position is accepted if there is
no overlap with either the solute or the remaining $n-1$ particles. The insertion probability is based on similar trial placements
averaged over $10^8-10^9$ insertions. For the present study involving solute and solvent of equal size, the radius of the 
coordination volume is the same as the cut-off radius of $r_c=1.1\sigma$, where $\sigma$ is the hard-sphere diameter.

     \subsection{Corrected cluster partition function}\label{sc:correction}
For the reference hard-sphere system, the reweighing approach was also used to obtain $p_0$, the probability of observing no particles in the inner shell of the  solute, and $\langle n\rangle$,  the average occupancy of the inner-shell of the solute. 
Then using the gas-phase cluster partition function ($\Xi^{(n)}$) and the relation $n! K_n^{(0)} = \Xi^{(n)}$, the activity 
correction $\zeta$ was obtained by solving for the roots of the polynomial equation corresponding to either Eq.~\ref{eq:22} or~\ref{eq:23}.

Including the correction, we then have 
  \begin{eqnarray}
\frac{\Delta c_n^{(0)}}{V} =  = \frac{1}{{n!}}\rho _0^{(s)}{\left( {X_A^{\left( p \right)}{\rho ^{\left( p \right)}}f_A^{\left( {s,p} \right)}\sqrt {{\kappa _{AA}}} } \right)^n}{\zeta ^n}{\Xi ^{(n)}}
     \label{eq:28}
    \end{eqnarray}
and the fraction of solute not bonded to any solvent molecule is
    \begin{equation}
    X_0^{\left( s \right)} = \frac{1}{{1 + \sum\limits_{n = 1}^{{n^{\max }}} {\frac{1}{{n!}}{{\left( {X_A^{\left( p \right)}{\rho ^{\left( p \right)}}f_A^{\left( {s,p} \right)}\sqrt {{\kappa _{AA}}} } \right)}^n}{\zeta ^n}{\Xi ^{(n)}}} }} \, .
     \label{eq:300}
    \end{equation}    
   The expressions of chemical potential for solute and solvent molecules with the corrected theory are given in the appendix.

\newpage
\section{Results and Discussions}\label{sc:results}
\subsection{Hard Sphere Reference} \label{sc:HSres}       
Table \ref{(table: Lagrange values)} gives the Lagrange multipliers ($\zeta$) corresponding to both the $n_{avg} = \langle n\rangle$ and
$p_0$ corrections. 
         \begin{table}[ht]
         	\caption{Lagrange multipliers for corrections based on $n_{avg}$ and $p_0$ for different reduced densities ($\rho \sigma^3$) }  
         	\begin{tabular}{|c| c| c|}
         		\hline\hline
         		$\rho \sigma^3$ & $n_{avg}$ (Eq.~\ref{eq:22}) & $p_0$ (Eq.~\ref{eq:23}) \\
         		\hline
         		0.2 &	1.386 &	1.383 \\
         		0.6 &	3.682 &	3.449 \\
         		0.7 &	5.232 &	4.838 \\
         		0.8	&   8.290 & 7.041 \\ 
         		0.9	&   15.646 & 11.173 \\
         		\hline
         	\end{tabular}
         	\label{(table: Lagrange values)}
         \end{table}     

\begin{figure*}[htp!]
      	\includegraphics[scale=1]{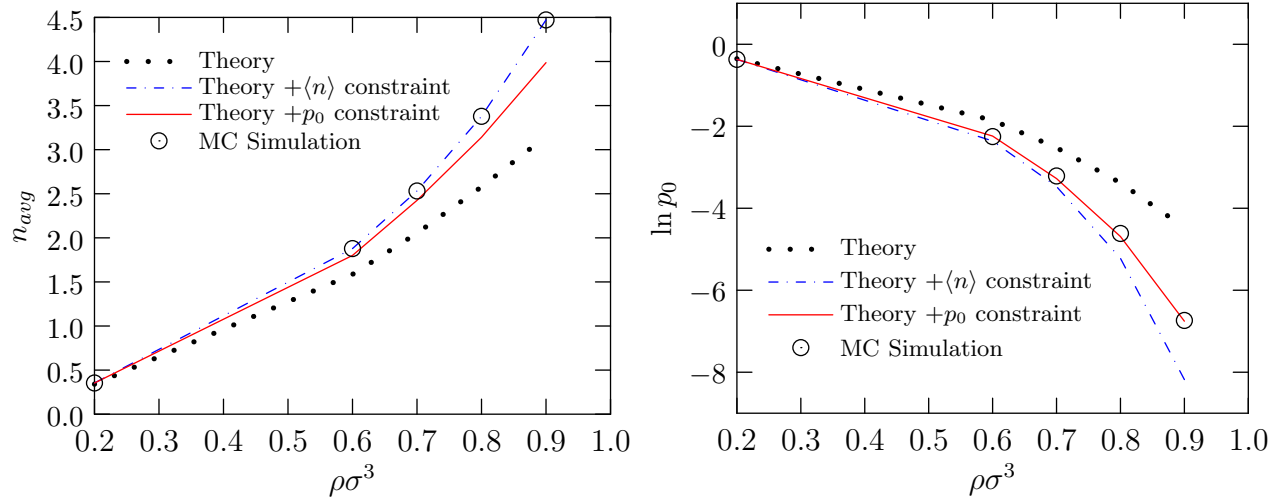}
 \caption{Comparison of  $n_{avg}$ (Eq.~\ref{eq:22}, left panel) and $p_0$ (Eq.~\ref{eq:23}, right panel)  for the packing of hard sphere reference fluid around a central hard-sphere solute.  ``Theory" indicates that only the gas-phase cluster partition function is used as in the original Marshall-Chapman approach \cite{marshall_molecular_2013}. }\label{fig:HScomparison}    
\end{figure*}

Fig.~\ref{fig:HScomparison} (Left panel) shows that the $n_{avg}$-based $\zeta$ captures the $n_{avg}$ across the density range, as it must since
the $\zeta$ is fit to reproduce this property.  
Likewise, Fig.~\ref{fig:HScomparison} (Right panel) shows that we can find factors $\zeta$ that
reproduce $p_0$ found in molecular simulations. In either of these case, it is evident that the gas-phase cluster partition function
can reproduce only the data at the lowest densities, emphasizing the limitations of the primitive quasichemical approach. 

\begin{figure}[h!]
\includegraphics[width=3.25in]{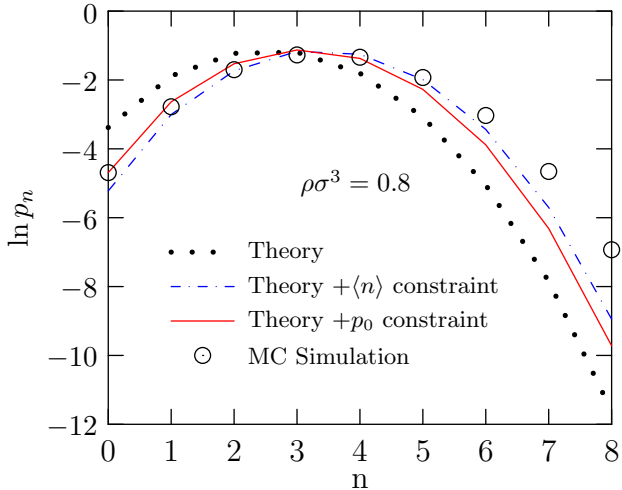}
 \caption{Coordination (occupancy) states in the inner-shell of the hard-sphere solute predicted for various approximations for the system density of $\rho\sigma^3 = 0.8$. Rest as in Fig.~\ref{fig:HScomparison}.}\label{fig:HScomparisonc}
\end{figure}

Fig.~\ref{fig:HScomparisonc} shows the entire occupancy distribution. First note that ignoring the bulk medium even the mode of the distribution is 
not correctly described. 
Thus it is not surprising that this approximation begins to hold only for densities below $\rho\sigma^3 = 0.2$ (Fig.~\ref{fig:HScomparison}). Both the
$n_{avg}$ and $p_0$ corrections lead to a better description of the low coordination states, with the $p_0$-based correction providing a better description of the
low-coordination data. This also highlights the importance of the low-coordination states in the free energy to populate the inner-shell of the solute, 
as was also found earlier for ions \cite{merchant_thermodynamically_2009}.

\subsection{Associating mixture}  \label{sc:Assores}
\begin{figure}[h!]
     	\includegraphics[scale=1]{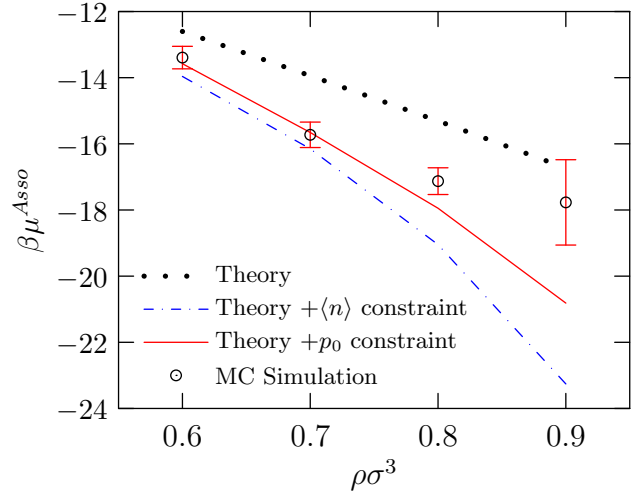}
     	\caption{Association contribution to chemical potential for a solute at different reduced densities. Solution is infinity dilute and energy of association between patchy-patchy and patchy-spherical molecules is 7~$k_BT$. The description of the labels is as in Fig.~\ref{fig:HScomparison}. }
     	\label{fig:chempotcomparison}
     \end{figure}

  Figure~\ref{fig:chempotcomparison} presents the central result of this study. 
 Notice that using just the
 gas-phase cluster partition function fails in reproducing the association contribution to the chemical potential of the solute ($\mu^{Asso}$) for $\rho\sigma^3 \geq 0.6$. But the trends suggest that the gas-phase approximation would be acceptable for low densities. Correcting the gas-phase cluster partition function using $\zeta$ (Eqs.~\ref{eq:22} and~\ref{eq:23}) leads to much better  agreement of the predicted $\mu^{Asso}$ with simulations for densities up to 0.8. Indeed, within the statistical uncertainties of the simulation,  the $p_0$-based $\zeta$ correction predicts $\mu^{Asso}$ of the solute up to $\rho\sigma^3 = 0.8$, consistent with expectations
 based on results noted in Fig.~\ref{fig:HScomparisonc} and the observed importance of low-coordination states in the thermodynamics of solvation\cite{merchant_thermodynamically_2009}. 

\begin{figure}[h!]
    	\includegraphics[scale=1]{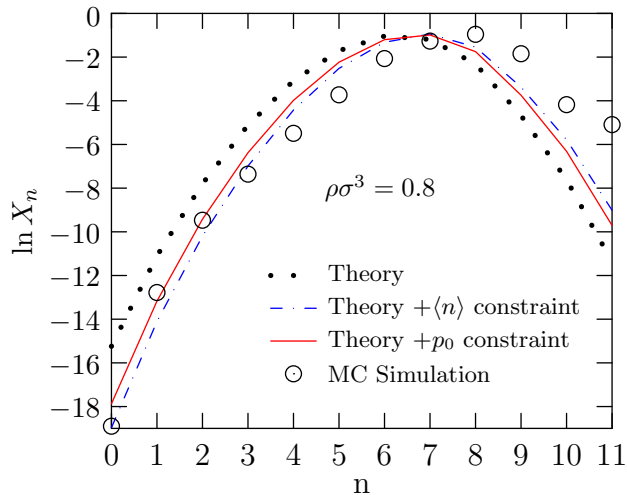}
    	\caption{Distribution of bonding states of solute for $\rho \sigma^3 = 0.8$. Rest as in Fig.~\ref{fig:chempotcomparison}. }
    	\label{fig:ASSOcomparison}    	
    \end{figure}

Fig.~\ref{fig:ASSOcomparison} shows the predicted bonding distribution.  We find that the $p_0$-based $\zeta$ correction better captures the low-$X_n$ states of the colloid. 
But the single-parameter correction of the gas-phase cluster partition function also has its limitations. In particular, the prediction of the high-bonding states is only qualitatively correct, but quantitatively it is not satisfactory. This discrepancy is partly due to the inability of the single-parameter correction  in capturing the high-coordination states in the reference (Fig.~\ref{fig:HScomparisonc}). One possible way to address this limitation,and this is part of our on-going research, is to include explicitly a molecular field to better describe the coordination shell population \cite{pratt_selfconsistent_2003} and 
address assumption Eq.~\ref{eq:yhs}.

\begin{figure}[h!]
    		\includegraphics[scale=1]{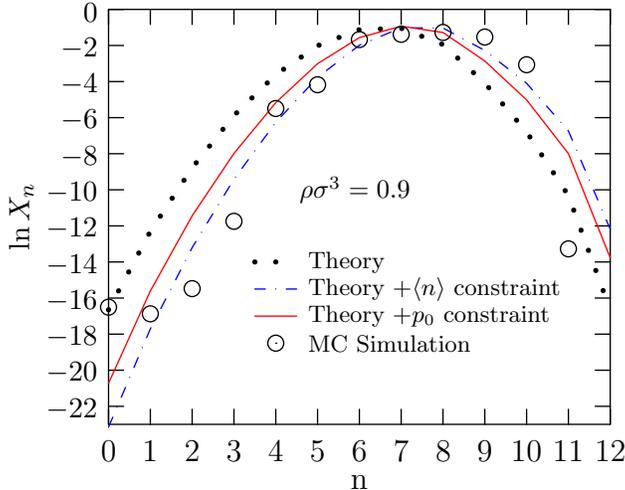}
      		\caption{Distribution of bonding states of solute for $\rho \sigma^3 = 0.9$. The nearly flat distribution near the mode and the dispersion in the data in the wings suggests problems in convergence for this high solvent density. Rest as in Fig.~\ref{fig:chempotcomparison}.}
      			\label{fig:rdhigh}
      	\end{figure}

For the highest density we considered, $\rho\sigma^3 = 0.9$, we observe a significant discrepancy between the predicted chemical potential and the value
obtained from simulations. For this case the statistical uncertainties in the simulated value were uncharacteristically high and characterizing the bonding state, even with the reweighed sampling approach, proved challenging. The system appears to have a nearly flat distribution of bonding states around the mode of the distribution, and the comparison between  theory ($p_0$-based $\zeta$) and simulations is also less than satisfactory (Fig.~\ref{fig:rdhigh}). 
Several test calculations also reveal severe system size limitations. For example, visual examination of configurations from a system with 256 particles
suggests the formation of chains of solvent molecules, akin to what might be expected in liquid crystals. Further  investigation of the high density state is required to better understand the discrepancy of theory predictions and simulations  for the $\mu^{Asso}$ at the reduced density of 0.9.

 \section{Conclusion}
            
   In this study we have developed a simple and effective way to model multi-body effects in colloidal mixtures. Building on Marshall and Chapman theory and borrowing ideas from quasichemical theory, we incorporate an improved representation of hard sphere reference fluid for better estimation of many body correlations. The key finding of our work is that, information about free energy to evacuate the observation shell around a solute in a hard sphere reference fluid can improve the estimation of bonding state of spherically symmetric colloids. We performed Monte Carlo simulations to test the theory and utilized ensemble reweighing approach to better reveal low bonding states with simulations.  Our comparative studies show a significant improvement with our approach over the Marshall Chapman theory for the bonding state and the excess chemical potential of the colloid, for the desired high density systems. In the next part of this study, a modified formulation with complete information from hard sphere will be presented and its effect on the association will be studied for various limiting cases.
   
   The present approach opens avenues to model a range of systems as a mixture of patchy and spherically symmetric colloids, with completely patchy and completely spherical being the extremes. The challenge in describing the multi-body effects can be handled by realizing the importance of packing in the reference system. For the current work, a symmetric mixture with equally sized patchy and spherically symmetric molecules with same strength of interaction was considered, extensions would be made for asymmetric mixtures with different sizes and association strength. This enables studies ranging from phase equilibria to study of new structures for complex systems with isotropic and anisotropic interactions.   
    
   \section{Acknowledgment} 
We thank Ben Marshall for helpful discussions.  We acknowledge RPSEA / DOE 10121-4204-01 and the Robert A. Welch Foundation (C-1241) for financial support

   \section{Appendix}
     The chemical potentials for solute $(\mu ^{AS(s)})$ and solvent $(\mu ^{AS(p)})$ with the corrected theory can be expressed as:    
     \begin{eqnarray}
     	\frac{{{\mu ^{AS(s)}}}}{{k_BT}} & = & \ln \left( {X_0^{\left( s \right)}} \right) - \frac{1}{2}\sum\limits_{A \in {\Gamma ^{\left( p \right)}}} {\left( {1 - X_A^{\left( p \right)}} \right){\rho ^{\left( p \right)}}} \frac{{\partial \ln {y_{HS}}\left( \sigma \right)}}{{\partial {\rho ^{\left( s \right)}}}} \nonumber \\
     	& + & \frac{{\overline n }}{2}{\rho ^{\left( s \right)}}\frac{{\partial \ln {y_{HS}}\left( \sigma \right)}}{{\partial {\rho ^{\left( s \right)}}}} - \overline n {\rho ^{\left( s \right)}}\frac{{\partial \ln \zeta }}{{\partial {\rho ^{\left( s \right)}}}}
     	\label{eq:310}
     \end{eqnarray}
     
      \begin{eqnarray}
      \frac{{{\mu ^{AS(p)}}}}{{k_BT}} & =  & \sum\limits_{A \in {\Gamma ^{\left( p \right)}}} {\ln \left( {X_A^{\left( p \right)}} \right)}+  \frac{{\overline n }}{2}{\rho ^{\left( s \right)}}\frac{{\partial \ln {y_{HS}}\left( \sigma \right)}}{{\partial {\rho ^{\left( p \right)}}}} \nonumber 
       \\ & & - \frac{1}{2}\sum\limits_{A \in {\Gamma ^{\left( p \right)}}} {\left( {1 - X_A^{\left( p \right)}} \right){\rho ^{\left( p \right)}}} \nonumber\frac{{\partial \ln {y_{HS}}\left( \sigma \right)}}{{\partial {\rho ^{\left( p \right)}}}} 
       \\ & &- \overline n {\rho ^{\left( s \right)}}\frac{{\partial \ln \zeta }}{{\partial {\rho ^{\left( p \right)}}}}
      \label{eq:32}
      \end{eqnarray}

\newpage

%
  
\end{document}